\documentclass[journal]{IEEEtran}

\usepackage{amsmath, amssymb}
\usepackage{graphicx}
\usepackage{booktabs}
\usepackage{multirow}
\usepackage{algorithm}
\usepackage{algorithmic}
\usepackage{tikz}
\usetikzlibrary{positioning,calc}
\usepackage{cite}
\usepackage{url}
\usepackage{xcolor}
\usepackage{listings}
\usepackage{xcolor}
\usepackage{placeins}
\usepackage{cuted}
\usepackage{capt-of}
\usepackage{booktabs}
\usepackage{graphicx}
\usepackage{tabularx}
\usepackage{multirow}
\usepackage{booktabs}
\usepackage{afterpage}
\newenvironment{agentprompt}
{\begin{quote}\itshape\small}
{\end{quote}}

\lstdefinestyle{jsonstyle}{
    basicstyle=\ttfamily\tiny,
    breaklines=true,
    breakatwhitespace=false,
    columns=flexible,
    frame=single,
    keepspaces=true,
    showstringspaces=false,
    tabsize=2,
    xleftmargin=0pt,
    xrightmargin=0pt,
    linewidth=\columnwidth
}
\title{An Explainable Agentic AI Framework for Distribution System Load Restoration}

\author{S. M. Monzurul Hoque Chowdhury and Qinghua Li%
\thanks{S. M. Monzurul Hoque Chowdhury and Qinghua Li are with the
Department of Electrical Engineering and Computer Science, University of
Arkansas, Fayetteville, AR, USA.}
\thanks{Corresponding author: Qinghua Li (email: qinghual@uark.edu).}
}

\begin{document}
\maketitle
\begin{abstract}
Distribution system load restoration is a sequential decision-making
problem in which restoration actions must maximize served load, prioritize
critical loads, and preserve power-flow feasibility. This paper proposes an
explainable agentic artificial intelligence (AI) framework for distribution
system load restoration. The proposed framework uses multiple large
language model (LLM) agents for planner guidance, strategic plan generation,
decision arbitration, memory-based feedback, and explanation generation.
Unlike unconstrained LLM-based control, the proposed method restricts agent
decisions to feasible restoration actions and evaluates all candidate plans
using deterministic AC power-flow feasibility checks. The framework is evaluated on IEEE 33-bus and 69-bus distribution feeders,
as well as a larger IEEE 118-bus test case, and compared against
reinforcement learning and planning baselines. The results show that the proposed agentic framework achieves
better restoration performance while producing structured explanations
for operator-facing decision support. A direct comparison with a
Single-LLM-Agent baseline further shows that the proposed multi-agent
decomposition improves weighted restored load and critical-load recovery
compared with a monolithic LLM decision structure.
\end{abstract}


\section{Introduction}
\label{sec:introduction}

Distribution system load restoration is a critical operational task after
faults, outages, or extreme weather events. The objective is to restore as
much load as possible, prioritize critical loads, and maintain stable network operation. This task is challenging because restoration requires
sequential decisions over load energization, feeder repair, tie-switch
operation, and distributed generation activation, while avoiding voltage
violations, line overloads, and power-flow infeasibility.

Existing restoration methods have made significant progress using
optimization, graph search, reinforcement learning, and hybrid
learning-control approaches \cite{pap01,pap04,pap11,pap17,pap18}.
Optimization and graph-search methods provide rigorous decision procedures,
while learning-based methods can adapt restoration policies from simulated
outage scenarios. However, learned policies usually provide limited
operator-facing explanation, and most optimization-based approaches require
detailed modeling assumptions or may become computationally expensive under
large outage scenarios. In practical restoration, operators need not only
recommended actions but also clear rationales for why those actions are
preferred over feasible alternatives.

Large language models (LLMs) create new opportunities for reasoning-based
decision support in power-system operation. Recent studies have explored
LLMs for power-system simulation, multi-agent solver co-ordination, autonomous research workflows,
and various representative tasks \cite{pap22,pap23,pap24,pap26}.
However, directly using an LLM as a grid controller is problematic because
unconstrained LLM outputs may include invalid actions, ignore physical
constraints, or produce decisions that are not guaranteed to satisfy feasibility requirements. In addition, prior work has noted that LLM-based
power-system applications may introduce reliability and security concerns, especially when generated outputs may affect critical
infrastructure decisions 
\cite{pap21}. These concerns motivate constrained LLM architectures that
combine language-based reasoning with feasible-action generation,
deterministic feasibility checking, fallback mechanisms, and transparent
decision explanations.

This paper proposes an explainable agentic AI framework for distribution
system load restoration. The framework decomposes restoration reasoning into
specialized LLM agents for planner guidance, strategic plan generation,
decision arbitration, memory-based feedback, and explanation generation. The
LLM agents are restricted to feasible restoration actions supplied by the
environment, and candidate plans are evaluated using deterministic AC
power-flow feasibility checks before execution. Thus, the framework uses
LLMs for plan-level reasoning and explanation while preserving
physics-aware validation in the decision loop.

The main contributions of this paper are:
\begin{itemize}
    \item A novel constrained agentic AI framework for distribution system load
    restoration that integrates feasible-action generation, topology-aware action shortlisting, deterministic AC power-flow plan
    evaluation, and memory feedback within a receding-horizon restoration
    loop.

    \item A safety-aware restoration decision loop in which LLM agents are
    not used as unconstrained grid controllers; all generated plans are
    restricted to feasible actions and verified using deterministic
    power-flow feasibility checks before execution.

    \item A specialized multi-agent LLM architecture that separates planner
    guidance, strategic plan generation, decision arbitration, memory
    feedback, and explanation generation, instead of relying on a single
    monolithic LLM controller.

    \item An operator-facing explanation mechanism that produces structured
    rationales for selected and rejected restoration plans using the same
    plan evaluations and feasibility indicators used during decision-making.

    \item Evaluation on IEEE 33-bus and 69-bus distribution feeders and a
    larger 118-bus restoration test case with randomized restoration
    scenarios, with comparisons against reinforcement learning, lookahead
    search, and Single-LLM-Agent baselines.
\end{itemize}

\section{Related Work}
\label{sec:related_work}

\subsection{Distribution System Restoration and Learning-Based Methods}
\label{subsec:restoration_rl}

Distribution system restoration has been widely studied using mathematical
optimization, graph search, rule-based methods, reinforcement learning, and
hybrid learning-control approaches. Graph reinforcement learning has been used for sequential distribution
system restoration \cite{pap01}, while reinforcement learning has also been applied to
critical-load restoration, microgrid-assisted restoration, resilient service restoration, and distribution network reconfiguration \cite{pap06,pap07,pap08,pap09, pap14, pap16}. Multi-agent
reinforcement learning is especially relevant because distribution systems
contain spatially distributed switches, loads, distributed generators, and
microgrids. Prior work has used multi-agent deep reinforcement learning along with
invalid-action masking or action space reduction to improve restoration and reconfiguration decisions
\cite{pap03,pap04,pap13}. These studies demonstrate the value of
learning-based and graph-aware restoration, but they focus mainly on learned
control policies rather than LLM-based planning, memory, and
operator-facing explanations.

\subsection{Optimization, Reconfiguration, and Hybrid Restoration}
\label{subsec:optimization_reconfiguration}

Optimization-based methods provide important
foundations for resilient distribution restoration. Robust optimization,
stochastic programming, and hybrid
learning-control methods have been used to improve restoration feasibility,
critical-load recovery, and operational resilience
\cite{pap11,pap17,pap18}. These methods motivate the
broader idea of combining high-level decision generation with deterministic
validation. However, many optimization and hybrid approaches are not
designed to provide structured natural-language rationales for selected and
rejected restoration plans, which limits their use as operator-facing
decision-support tools.

\subsection{LLM Agents and Agentic AI for Power Systems}
\label{subsec:llm_power_systems}

LLMs have recently been explored for power-system simulation, analysis,
optimization, graph-based reasoning, autonomous research workflows, and
multi-agent solver coordination \cite{pap22,pap23,pap24,pap25,pap26,pap27}.
These studies suggest that LLMs can support power-system tasks when
connected to external tools and deterministic solvers. More broadly, LLM-agent frameworks such as ReAct, AutoGen, CAMEL,
Reflexion, Toolformer, and HuggingGPT show how language models can support
reasoning, multi-agent coordination, tool use, feedback, and task
orchestration
\cite{pap30,pap31,pap32,pap34,pap36,pap37}. However, none of these works address distribution system load restoration under feasible-action
constraints and AC power-flow feasibility requirements.

\subsection{Research Gap}
\label{subsec:research_gap}

Existing restoration studies have made strong progress in optimization-based
restoration, graph reinforcement learning, multi-agent reinforcement
learning, and hybrid learning-control
methods. In parallel, recent LLM-agent studies show that language models can
coordinate with tools and deterministic solvers. However, there remains a
gap in constrained, explainable, agentic AI for distribution system load
restoration.

To the best of our knowledge, this paper presents the first constrained agentic LLM framework for distribution
system load restoration that jointly integrates feasible-action generation,  topology-aware action shortlisting, multi-step restoration
planning,
deterministic AC power-flow validation, memory feedback, and operator-facing
explanation. The proposed framework embeds specialized LLM agents inside a
physics-aware restoration loop, where language-based reasoning is used for
plan generation, decision arbitration, and explanation, while executable
restoration actions remain constrained by the feasible action set and
deterministic feasibility checks.

\section{Methodology}
\label{sec:methodology}

This paper proposes an explainable agentic AI framework for distribution
system load restoration. The objective is to determine a sequence of
restoration actions that maximizes priority-weighted restored load,
prioritizes critical loads, and maintains stable power-system operation.
The proposed method does not use LLMs as
unconstrained grid controllers. Instead, the LLM agents operate inside a
constrained restoration loop in which action validity, candidate-action
screening, plan-level reasoning, and physical feasibility verification are
handled by separate components.

This separation is important because these components require different
types of decision logic. Action validity is a topology-based check that
determines which restoration actions are available in the current grid
state. Candidate-action screening ranks the valid actions according to
restoration value, future reachability, and critical-load priority. Plan-level reasoning then uses the shortlisted actions to
construct and compare multi-step restoration strategies, while deterministic
AC power-flow evaluation verifies whether the resulting network
satisfies operating constraints. By separating these roles, the framework
allows the LLM agents to focus on high-level restoration reasoning and
explanation, while deterministic modules enforce action validity and
physical feasibility.

\begin{figure*}[t]
\centering
\begin{tikzpicture}[
    block/.style={
        rectangle,
        draw,
        rounded corners,
        align=center,
        minimum width=2.8cm,
        minimum height=0.9cm,
        font=\small
    },
    arrow/.style={->, thick},
    feedback/.style={->, thick, dashed},
    node distance=1.15cm and 1.05cm
]

\node[block] (state) {Current grid\\state $s_t$};

\node[block, right=of state] (legal) {Feasible action\\generation};

\node[block, right=of legal] (shortlist) {Topology-aware\\action shortlisting};

\node[block, right=of shortlist] (llm) {LLM agentic\\planning};

\node[block, below=1.65cm of llm] (eval) {AC power-flow\\plan evaluation};

\node[block, left=of eval] (execute) {Decision and execute\\first action};

\node[block, below=1.65cm of legal] (update) {State + memory\\update};

\node[block, left=of update] (explain) {Explainability\\output};

\draw[arrow] (state.east) -- (legal.west);
\draw[arrow] (legal.east) -- (shortlist.west);
\draw[arrow] (shortlist.east) -- (llm.west);
\draw[arrow] (llm.south) -- (eval.north);
\draw[arrow] (eval.west) -- (execute.east);
\draw[arrow] (execute.west) -- (update.east);
\draw[arrow] (update.west) -- (explain.east);

\draw[feedback] (update.north) -- (legal.south);

\node[font=\scriptsize, anchor=east]
    at ($(update.north)!0.5!(legal.south)+(-0.10cm,0)$)
    {receding-horizon loop};

\coordinate (memA) at ($(update.north east)+(0.25cm,0)$);
\coordinate (memB) at ($(update.north east)+(0.25cm,0.85cm)$);
\coordinate (memC) at ([xshift=-0.45cm]llm.south |- memB);
\coordinate (memD) at ([xshift=-0.45cm]llm.south);

\draw[feedback]
    (update.north east) -- (memA)
    -- (memB)
    -- node[midway, above, font=\scriptsize] {memory feedback}
    (memC)
    -- (memD);

\end{tikzpicture}
\caption{Overview of the proposed agentic restoration framework. Feasible
actions are shortlisted, planned over by LLM agents, verified using AC
power-flow checks, and executed in a receding-horizon loop.}
\label{fig:framework_overview}
\end{figure*}
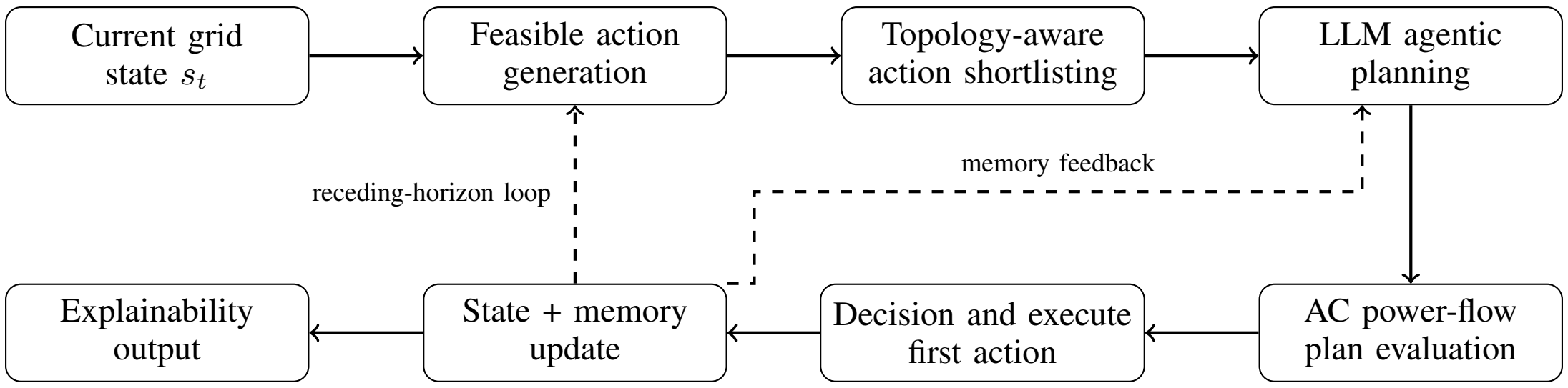




\subsection{Restoration Problem Formulation}
\label{subsec:problem_formulation}

The restoration task is formulated as a finite-horizon sequential
decision-making problem. At each restoration step $t$, the agentic framework observes
the current distribution-grid state $s_t$ and selects one restoration action
$a_t$ from the feasible action set $\mathcal{A}(s_t)$. From a theoretical sequential decision-making perspective, the restoration objective can be expressed as maximizing cumulative restoration reward over a finite horizon:
\begin{equation}
    \max_{\pi} \sum_{t=0}^{T-1} r(s_t,a_t),
\end{equation}
where $T$ is the restoration horizon, $\pi$ is the restoration policy, and
$r(s_t,a_t)$ is the immediate reward obtained after applying action $a_t$ in
state $s_t$.

The selected action $a_t$ must satisfy:
\begin{equation}
    a_t \in \mathcal{A}(s_t),
\end{equation}
where $\mathcal{A}(s_t)$ is constructed from the current network topology,
line status, load status, tie-switch status, reachable buses, and
distributed-generator status. This feasible-action constraint is central to the
proposed framework because it prevents the LLM agents from inventing
restoration actions that are not feasible in the current grid state.

The restoration action space contains four executable action categories:
\begin{equation}
a_t \in \{
\text{serve\_load},
\text{repair\_line},
\text{switch\_tie},
\text{start\_dg}
\}.
\end{equation}
The action \textbf{serve\_load} energizes a currently disconnected load that
is reachable from an active source. The action \textbf{repair\_line}
restores a failed feeder segment. The action \textbf{switch\_tie} closes a
normally open tie line to create an alternative restoration path. The action
\textbf{start\_dg} activates a distributed generator as an additional source.

After each action, AC power flow is executed. The resulting network state
must satisfy:
\begin{equation}
    V_{\min} \leq V_i \leq V_{\max}, \quad \forall i \in \mathcal{N},
\end{equation}
\begin{equation}
    \ell_j \leq \ell_{\max}, \quad \forall j \in \mathcal{L},
\end{equation}
where $V_i$ is the voltage magnitude at bus $i$, $\ell_j$ is the loading
percentage of line $j$, and $\mathcal{N}$ and $\mathcal{L}$ denote the bus
and line sets, respectively. If AC power flow does not converge, or if
voltage or line-loading constraints are violated, the action is treated as
infeasible and the previous network state is restored.
\subsection{Scenario-Based Restoration Environment}
\label{subsec:restoration_environment}

For the IEEE 33-bus and 69-bus distribution feeders and the larger 118-bus
restoration test case, the base network topology is fixed, while the
restoration scenario varies across random seeds. Each scenario defines
the load demand values, priority weights, critical-load locations, failed
feeder sections, and initial outage condition. Therefore, the evaluation is performed by
measuring restoration performance across multiple outage and loading
conditions on the same feeder template, rather than claiming generalization
to arbitrary unseen feeder topologies.

The environment state includes the current restoration step,
active source buses, started distributed generators, reachable
buses, failed feeder sections, open tie
lines, restored loads, restored critical loads, weighted restored
load, and the number of remaining feasible actions.
\subsection{Agentic Restoration Architecture}
\label{subsec:agentic_architecture}

Fig.~\ref{fig:framework_overview} shows the overall agentic restoration framework.
At each restoration step, the environment first constructs the feasible
action set from the current grid state and applies topology-aware action shortlisting. The shortlisted actions are provided to
the LLM agents, which generate candidate multi-step restoration plans,
arbitrate among evaluated plans, store recent outcomes, and explain the
selected action. Before execution, candidate plans are simulated in a cloned
environment and evaluated using deterministic AC power flow. Only the first
action of the selected plan is executed, after which the grid state is
updated and the process repeats.

The proposed framework decomposes the restoration process into specialized
LLM agents rather than relying on a single monolithic LLM controller. 
\begin{itemize}
    \item \textbf{\textit{Planner Guidance Agent}}: provides high-level rollout
    preferences, including preferred action types, action types to avoid,
    priority bus focus, and planning horizon.

    \item \textbf{\textit{Strategic Planning Agent}}: generates multiple candidate
    multi-step restoration plans using only the candidate actions supplied
    by the environment.

    \item \textbf{\textit{Decision Agent}}: arbitrates among evaluated candidate
    plans and selects the final plan index.

    \item \textbf{\textit{Memory Agent}:} stores recent action outcomes and provides
    feedback about successful actions, failed actions, risky patterns, and
    useful action sequences.

    \item \textbf{\textit{Explainability Agent}:} generates a structured explanation
    of the selected plan, selected first action, and rejected alternatives.
\end{itemize}

The LLM agents are used for reasoning, planning, arbitration, and
explanation. They do not directly enforce physical feasibility. Feasibility
is enforced by the deterministic network-update module, which applies
restoration actions, runs AC power-flow checks, and rejects infeasible
transitions.

All the LLM agents receive a shared task description that defines the restoration setting, available action types, system objective, and physical feasibility requirements. 
\begin{agentprompt}
You are helping solve a power distribution service restoration problem. The
available action types are \textbf{serve\_load}, \textbf{repair\_line},
\textbf{switch\_tie}, and \textbf{start\_dg}. The system objective is to
maximize restored weighted load, prioritize critical loads, maintain AC
power-flow feasibility, and avoid unsafe or unstable choices. AC power flow
must converge, all bus voltages must remain within the specified limits, and
all line loadings must remain below the maximum loading threshold. You must
choose only from provided actions or provided plans. Do not invent buses,
lines, switches, generators, or actions.
\end{agentprompt}

This shared instruction is designed to reduce hallucination and keep the LLM agents grounded in the restoration task.
It is foundational to the proposed design. The LLM agents are not allowed to freely generate arbitrary restoration actions. Instead, they must
operate over a restricted set of valid candidates supplied by the
environment and screened by the topology-aware action shortlist generator.

\subsection{Feasible Action Generation and Topology-Aware
Action Shortlisting}
\label{subsec:feasible_action_shortlisting}

At each restoration step, the framework first constructs the feasible action
set $\mathcal{A}(s_t)$ from the current grid state. This step is performed
before any LLM agent is called. A load-restoration action is feasible only if
the load bus is currently reachable from an active source and the load has
not already been restored. A line-repair action is feasible only if the
corresponding backbone feeder section is currently failed. A tie-switch
action is feasible only if the corresponding tie line is currently open. A
DG-start action is feasible only if the candidate distributed generator has
not already been activated.

The feasible action set acts as a hard constraint on LLM-based decision
generation. The LLM agents do not have the freedom to choose from the full
combinatorial action space and are not allowed to generate arbitrary bus,
line, switch, or generator actions. Instead, all downstream planning is
restricted to action strings supplied by the environment. If a proposed
action does not belong to the current feasible action set, it is rejected as
invalid.

After the feasible action set is constructed, the actions are passed to a
topology-aware action shortlist generator. This module screens
the feasible actions and suggests a compact set of promising candidate
actions $\mathcal{A}_c(s_t)$ for the LLM agents. The shortlist generator does
not create new actions and does not make the final restoration decision. It
identifies promising actions from the feasible action set using
priority-weighted restoration value, critical-load priority, and graph
reachability.

For a feasible action $a \in \mathcal{A}(s_t)$, the internal shortlisting
score is computed as:
\begin{equation}
    S(a)
    =
    \alpha_1 O(a)
    +
    \alpha_2 U(a)
    +
    \alpha_3 C(a),
\end{equation}
where $O(a)$ is the immediate priority-weighted load restoration opportunity,
$U(a)$ is the future unlock term, and $C(a)$ is the critical-load priority
term. The fixed weights used in this study are
$\alpha_1=0.40$, $\alpha_2=0.25$, and $\alpha_3=0.35$. These values emphasize
immediate priority-weighted restoration and critical-load recovery, while
still accounting for future reachability.

For a load-restoration action at bus $b$, the opportunity term is computed as
\begin{equation}
    O(a)=P_b w_b,
\end{equation}
where $P_b$ is the active load demand and $w_b$ is the scenario-specific
priority weight of bus $b$. The term $C(a)$ is
determined by the scenario-specific critical-priority assignment of bus $b$.
In the experiments, $C(a)$ is set to $1$ for actions that restore a critical
load and $0$ otherwise.

A bus is considered reachable if it has an energized path from an active
source bus or an activated distributed generator to that bus through
available feeder sections and closed switches. The future
unlock term $U(a)$ is computed as the increase in reachable load after
applying action $a$ in a cloned environment:
\begin{equation}
    U(a)=L_{\mathrm{reach}}(s_t,a)-L_{\mathrm{reach}}(s_t),
\end{equation}
where $L_{\mathrm{reach}}(s_t)$ is the total reachable load before applying
$a$, and $L_{\mathrm{reach}}(s_t,a)$ is the total reachable load after
applying $a$. The values of $O(a)$ and $U(a)$ are measured in load units,
while $C(a)$ is a binary critical-load indicator. Therefore, the
shortlisting score is used only as a relative ranking score over the current
feasible action set.

The output of this module is the shortlisted candidate action set
$\mathcal{A}_c(s_t)$. The LLM agents then take this candidate action set as
input and generate diverse multi-step restoration plans using only these
shortlisted feasible actions. AC power-flow feasibility and risk are not
evaluated during this shortlisting step. Instead, after the candidate plans
are generated, each plan is simulated in a cloned environment and evaluated
using deterministic AC power flow before the final action is selected for
execution. Thus, the
shortlist generator provides topology-aware local action
screening, while physics-aware feasibility and risk evaluation are enforced
later through deterministic candidate-plan evaluation. This separation
prevents the LLM from being used as a brute-force search mechanism over all
possible restoration actions while ensuring that executed actions are
selected only after physics-based feasibility checks.

\subsection{Planner Guidance Agent}
\label{subsec:planner_guidance}

The Planner Guidance Agent observes the current state, candidate action set,
and memory summary. Its role is not to select the final restoration action.
Instead, it returns structured guidance that influences short-horizon
rollouts during plan evaluation. The output includes:
\begin{itemize}
    \item \textbf{preferred\_action\_types}: action types encouraged during
    rollout;
    \item \textbf{avoid\_action\_types}: action types discouraged during
    rollout;
    \item \textbf{priority\_bus\_focus}: buses receiving additional
    restoration focus;
    \item \textbf{planning\_horizon}: the rollout depth used for
    short-horizon evaluation.
\end{itemize}

The prompt used by the Planner Guidance Agent is summarized as follows:
\begin{agentprompt}
You are the planner-guidance agent. Your job is not to choose the final
action. Your job is to guide how short-horizon rollouts should behave.
Return strict JSON containing preferred action types, avoided action types,
priority bus focus, and planning horizon. The planning horizon must be an
integer between 1 and 6. Preferred and avoided action types can only be
\textbf{serve\_load}, \textbf{repair\_line}, \textbf{switch\_tie}, and
\textbf{start\_dg}. Choose only bus IDs that appear relevant from the
provided state. Do not include extra text.
\end{agentprompt}

This guidance is used only as a soft bias during short-horizon rollout. It
does not override the feasible action set or deterministic AC power-flow
feasibility checks.
\subsection{Strategic Planning Agent}
\label{subsec:strategic_planning}

The Strategic Planning Agent generates a diverse set of candidate
restoration plans. Each plan is a short sequence of shortlisted valid
actions:
\begin{equation}
    p_k = \{a_{k,1}, a_{k,2}, \ldots, a_{k,H}\},
\end{equation}
where $p_k$ is the $k$-th candidate plan and $H$ is the plan horizon.

The role of the Strategic Planning Agent is not to rescore individual
actions already screened by the topology-aware action shortlist generator.
Instead, it constructs multiple candidate restoration strategies from the
shortlisted valid action set. For example, given the same candidate actions,
the agent may propose a load-first plan, a repair-first plan, a tie-switching
plan, or a DG-supported plan. These plans represent different hypotheses
about how near-term actions may affect future restoration opportunities.

The prompt used by the Strategic Planning Agent is summarized as follows:
\begin{agentprompt}
You are the main strategic planning agent. Generate a diverse set of
plausible multi-step restoration plans. Return strict JSON with a list of
plans. Generate four to six plans. Each plan length must be three to six.
Every action must come exactly from the provided candidate actions. Make
plans structurally diverse, including load-first, repair-first, tie-first,
and DG-first plans when meaningful. Prefer plans that restore high-priority
or critical load. Do not invent actions. Do not include extra text.
\end{agentprompt}

This stage uses the LLM for strategic diversity and plan-level reasoning,
but the LLM is still constrained by the candidate action set. Therefore,
plan generation is not an unconstrained text-generation problem; it is a
structured planning step over a verified set of valid restoration actions.
\subsection{Physics-aware Plan Evaluation}
\label{subsec:plan_evaluation}

Each candidate plan is evaluated by a deterministic physics-aware evaluator.
The evaluator clones the current restoration environment and executes the
candidate plan step by step in the cloned environment. After each action,
AC power flow is run. If the power flow fails to converge, or if voltage or
line-loading constraints are violated, the plan is marked infeasible and
penalized.

For each candidate plan $p_k$, the evaluator records feasibility status,
restored load, restored critical-load count,
weighted restored load, voltage or line-loading violations, and
the number of executed steps. These quantities are returned as structured
plan-evaluation evidence for the Decision Agent. The Decision Agent therefore
does not rely on a single scalar score alone; instead, it selects among the
evaluated candidate plans by considering feasibility,
critical-load recovery, weighted restored load, and memory context.
\subsection{Decision Agent and Receding-Horizon Execution}
\label{subsec:decision_agent}

After all candidate plans are evaluated, the Decision Agent receives the
current state, candidate plans, deterministic plan evaluations, and memory
summary. It then selects the best plan among the evaluated candidates. 

The prompt used by the Decision Agent is summarized as follows:
\begin{agentprompt}
You are the final decision agent. You must choose the best plan index from
the provided evaluated plans. Return strict JSON with the selected plan
index and confidence. Prefer feasible plans, more critical
restoration, and stronger weighted restoration. Do not include extra text.
\end{agentprompt}

Once the final plan $p_{k^*}$ is selected, the framework executes only the
first action of that plan:
\begin{equation}
    a_t = p_{k^*}(1).
\end{equation}

Only the first action is executed because the framework follows a
receding-horizon restoration strategy. After one action is applied, the
grid state, reachable buses, feasible action set, and AC power-flow feasibility
conditions may change. Therefore, the remaining actions in the selected
plan are used as lookahead reasoning rather than committed actions, and the
framework replans from the updated state at the next restoration step. 

The environment then runs AC power flow, updates the grid state, records the
action outcome, and repeats the planning process at the next step. This
creates a receding-horizon restoration process: the LLM agents reason over
multi-step plans, but only one verified action is executed before the state is observed again.
\vspace{-0.3em}
\subsection{Memory Agent}
\label{subsec:memory_agent}

The Memory Agent stores recent restoration outcomes within each episode.
For each executed action, it records the action string,
feasibility status, failure reason, restored load, restored critical-load
count, weighted restored load, and selected plan signature.

The memory summary includes:
\begin{itemize}
    \item recently successful action types;
    \item recently failed action types;
    \item risky action patterns;
    \item good action sequences;
    \item average, best, and worst recent plans.
\end{itemize}

This memory summary is provided to all the other four LLM agents. Therefore, the framework can
favor action types or action patterns that recently performed well and
discourage actions associated with infeasibility or poor performance.
\subsection{Explainability Agent}
\label{subsec:explainability_agent}

After the final plan is selected, the Explainability Agent generates a
structured decision explanation. It receives the current state, candidate
plans, deterministic plan evaluations, selected plan, and memory
summary. The explanation includes:
\begin{itemize}
    \item a concise summary of the decision context;
    \item descriptions of the candidate plans;
    \item the selected plan and selected first action;
    \item reasons for selecting the chosen plan;
    \item reasons for rejecting alternative plans.
\end{itemize}

The prompt used by the Explainability Agent is summarized as follows:
\begin{agentprompt}
You are the explainability agent. Explain what candidate plans were
suggested, why the chosen plan/action was selected, and why the other plans
were not selected. Return strict JSON containing a summary, plan
explanations, and decision explanation. Use only the provided state,
candidate plans, plan evaluations, selected plan index, and memory summary.
Do not include extra text.
\end{agentprompt}

Since the explanation is generated from the evaluated plans, feasibility
flags, restored-load metrics, selected plan index, and memory
summary, it is grounded in the actual decision artifacts used by the framework. The explanation is therefore not a detached post-hoc narrative;
it summarizes the same plan evaluations and constraints used during decision-making.
\vspace{-0.2em}
\subsection{Structured Output Validation and Fallback}
\label{subsec:structured_output}

All LLM agents are instructed to return structured JSON outputs. The
framework parses each LLM response and validates the required fields. If an
LLM call fails, returns no text, produces invalid JSON, selects an invalid
index, or generates actions outside the candidate action set, the framework
uses a deterministic fallback. 

For plan generation, fallback plans are constructed from the candidate action
set. For decision selection, if the Decision Agent output is invalid or
missing, the fallback chooses the candidate plan with the highest deterministic
fallback value score. For each candidate plan $p_k$, this score is computed as
\begin{equation}
\begin{aligned}
S_{\mathrm{fb}}(p_k)
&=
\lambda_1 L_{\mathrm{w}}(p_k)
+\lambda_2 N_{\mathrm{crit}}(p_k) \\
&\quad
-\lambda_3 V_{\mathrm{viol}}(p_k)
-\lambda_4 I_{\mathrm{viol}}(p_k).
\end{aligned}
\end{equation}
where $L_{\mathrm{w}}(p_k)$ is the weighted restored load after executing
plan $p_k$, $N_{\mathrm{crit}}(p_k)$ is the number of restored critical
loads, $V_{\mathrm{viol}}(p_k)$ denotes voltage-limit violations, and
$I_{\mathrm{viol}}(p_k)$ denotes line-loading violations. The corresponding coefficients represent configurable weights that can be adjusted according to operator-defined restoration priorities and current system operating conditions. For instance, critical-load restoration may receive greater emphasis during emergency scenarios, while voltage and line-loading penalties may be assigned greater importance under stressed network conditions. 

This validation-and-fallback mechanism improves reliability by preventing
malformed LLM outputs from directly affecting the restoration execution. The
final executed action must still pass the deterministic network-update
procedure and AC power-flow feasibility checks.
\subsection{Overall Restoration Algorithm}
\label{subsec:algorithm}

Algorithm~\ref{alg:agentic_restoration} summarizes the proposed explainable
agentic AI restoration procedure.

\begin{algorithm}[t]
\caption{Explainable Agentic AI Framework for Load Restoration}
\label{alg:agentic_restoration}
\begin{algorithmic}[1]
\STATE Initialize restoration environment from scenario seed
\STATE Initialize Planner Guidance, Strategic Planning, Decision, Memory,
and Explainability Agents
\WHILE{maximum step limit is not reached and feasible actions remain}
    \STATE Observe current grid state $s_t$
    \STATE Construct feasible action set $\mathcal{A}(s_t)$
    \STATE Feed  $\mathcal{A}(s_t)$ into the topology-aware action shortlist generator
    \STATE Obtain candidate action set $\mathcal{A}_c(s_t)$
    \STATE Planner Guidance Agent generates rollout guidance
    \STATE Strategic Planning Agent generates candidate plans
    $\{p_1,\ldots,p_K\}$ using $\mathcal{A}_c(s_t)$
    \FOR{each candidate plan $p_k$}
        \STATE Simulate plan in a cloned restoration environment
        \STATE Run AC power flow after each action
        \STATE Record feasibility, restored load, critical-load
        recovery, weighted restored load, and constraint violations
    \ENDFOR
    \STATE Decision Agent selects the best plan using the evaluated
    candidate-plan evidence and memory summary
    \STATE Execute the first action of the selected plan
    \STATE Run AC power flow and update the grid state
    \STATE Update Memory Agent with the action outcome
    \STATE Explainability Agent generates the decision explanation
\ENDWHILE
\STATE Return final restoration metrics
\end{algorithmic}
\end{algorithm}
\section{Experimental Setup}
\label{sec:experimental_setup}
\subsection{Test Environments and Scenario Generation}
\label{subsec:environment}

The proposed framework is evaluated on IEEE 33-bus and 69-bus distribution
feeders, together with a larger 118-bus restoration test case. The 33-bus
and 69-bus systems are used as distribution restoration benchmarks, while
the 118-bus case is included to examine scalability on a larger network.
All environments are implemented in \textbf{pandapower}. For each system,
the base feeder structure is fixed, while restoration scenarios vary across
random seeds through load demand values, priority weights, critical-load
locations, failed feeder sections, and initial restoration conditions. Here,
load-priority weights denote the relative restoration importance used in the
weighted-restoration objective. Critical-load assignments identify loads
that represent essential service categories; these loads are assigned the
highest priority weights and are also counted separately in the critical-load
recovery metric.

\begin{table}[t]
\centering
\caption{Summary of restoration test cases.}
\label{tab:test_cases}
\scriptsize
\setlength{\tabcolsep}{3pt}
\renewcommand{\arraystretch}{1.05}
\begin{tabular}{@{}lccp{0.43\columnwidth}@{}}
\hline
System & Buses & Action Types & Scenario Variability \\
\hline
IEEE 33  & 33  & 4 & Load values, priorities, critical loads, outages \\
IEEE 69  & 69  & 4 & Load values, priorities, critical loads, outages \\
IEEE 118 & 118 & 4 & Load values, priorities, critical loads, outages \\
\hline
\end{tabular}
\end{table}

For each system, 250 restoration scenarios are generated and divided into
200 training scenarios, 25 validation scenarios, and 25 held-out test
scenarios. The training and validation scenarios are used only for the
learning-based baselines. Specifically, the reinforcement-learning baselines
are trained on the training scenarios, while the validation scenarios are
used for checkpoint and hyperparameter selection. The proposed agentic LLM
framework does not require task-specific training or validation. Final
performance for all methods is reported on the same 25 held-out test
scenarios to ensure a paired and fair comparison. For each held-out scenario,
all methods are initialized with the same feeder condition, failed lines,
load values, priority weights, critical buses, and DG/tie-switch
configuration. This paired
evaluation protocol ensures that performance differences are due to the
restoration strategy rather than different test conditions.

Every candidate action is checked using AC power flow. The feasibility
criteria are:
\begin{equation}
    0.95 \leq V_i \leq 1.05, \quad \forall i,
\end{equation}
and
\begin{equation}
    \ell_j \leq 100\%, \quad \forall j.
\end{equation}
If the resulting state violates AC power-flow feasibility, the attempted
action is canceled, the network is reset to the feasible state that existed
before the action was applied, and a penalty is assigned. The restoration
loop then continues from this feasible state, and subsequent decisions are
made from the remaining valid candidate actions.
\subsection{Evaluation Metrics}
\label{subsec:evaluation_metrics}

The restoration performance is evaluated using two priority-aware metrics:
weighted restored load and restored critical-load. Weighted restored load
captures the amount of restored load after accounting for heterogeneous
priority weights, while critical-load measures the amount of restored
demand associated with critical loads.

The weighted restored load is defined as:
\begin{equation}
    L_{\mathrm{weighted}}
    =
    \sum_{b \in \mathcal{B}_{\mathrm{served}}} P_b w_b,
\end{equation}
where $P_b$ is the active load demand at bus $b$, $w_b$ is its priority
weight, and $\mathcal{B}_{\mathrm{served}}$ is the set of restored load
buses.

The restored critical-load is defined as:
\begin{equation}
    P_{\mathrm{crit}}
    =
    \sum_{b \in \mathcal{B}_{\mathrm{served}} \cap \mathcal{B}_{\mathrm{crit}}}
    P_b,
\end{equation}
where $\mathcal{B}_{\mathrm{crit}}$ is the set of critical-load buses. These
two metrics are used together because weighted restored load measures the
overall priority-aware restoration objective, while restored critical-load measures recovery of essential service loads.
\subsection{LLM Configuration}
\label{subsec:llm_config}

The LLM agents are implemented using GPT-5.4 through the OpenAI Responses API, with the same model configuration used across all agent calls and experimental scenarios. All LLM agents are prompted to
return strict JSON outputs. The framework includes JSON extraction, parsing,
and validation to ensure that the agents' outputs can be used safely by the restoration pipeline.
\subsection{Baselines}
\label{subsec:baselines}

The proposed framework is compared against representative reinforcement
learning, lookahead search, and LLM-based baselines. The reinforcement
learning baselines are \textbf{Deep Q-Network (DQN-RL)} and \textbf{ActorCritic-RL}. The
lookahead search baseline is \textbf{Monte Carlo Tree Search (MCTS)-Rollout}. The LLM-based ablation
baseline is \textbf{SingleLLMAgent}. The reinforcement learning baselines
evaluate whether policies trained from restoration scenarios can learn
effective sequential restoration behavior. The lookahead search baseline
evaluates whether simulated lookahead over the same restoration environment
can identify strong actions without LLM-based plan generation. The LLM-based
ablation evaluates whether the proposed multi-agent decomposition provides
benefits beyond a monolithic LLM controller.

The DQN baseline learns an action-value function $Q(s,a)$ over the
restoration environment using the training scenarios. At each decision step,
the trained DQN selects a restoration action by estimating the expected
long-term return of each available action. The Actor--Critic baseline learns
two components: an actor network that selects restoration actions and a
critic network that estimates the value of the current restoration state. The MCTS-Rollout baseline performs
fixed-budget tree search over candidate restoration actions. At each decision
step, MCTS expands promising actions, performs fixed rollouts, and chooses
the action with the highest estimated value.

The SingleLLMAgent baseline is included to isolate the effect of agentic
decomposition. This baseline receives the same restoration environment,
shortlisted candidate action set, memory summary, and deterministic
physics-aware plan evaluation as the proposed framework.
However, instead of assigning planner guidance, strategic plan generation,
decision arbitration, and explanation to separate agents, it performs these
LLM-based reasoning tasks within one LLM call. The agent is still required to
use only the provided candidate actions, and its generated plans are
validated before execution.
\subsection{Restoration Horizon}
\label{subsec:restoration_horizon}

The restoration process is evaluated under a finite action budget. Let
$T_{\max}$ denote the maximum number of restoration steps allowed in one
episode. At each step, the controller selects one feasible
action, such as serving a load, repairing a line, closing a tie switch, or
activating a distributed generator. Therefore, $T_{\max}$ limits the total
number of corrective actions that can be executed during one restoration
episode.

A smaller value of $T_{\max}$ represents a more restrictive restoration
setting, while a larger value allows to execute additional
restoration actions. In this study, all methods are evaluated using
$T_{\max}=10$ and $T_{\max}=15$ to compare performance under shorter and
longer restoration horizons.

\section{Results and Discussion}
\label{sec:results}

\begin{table}[t]
\centering
\caption{IEEE 33-bus restoration results.}
\label{tab:ieee33_results}
\footnotesize
\setlength{\tabcolsep}{4pt}
\renewcommand{\arraystretch}{1.10}
\begin{tabularx}{\columnwidth}{@{}Xcccc@{}}
\toprule
\multirow{2}{*}{Method} &
\multicolumn{2}{c}{$T_{\max}=10$} &
\multicolumn{2}{c}{$T_{\max}=15$} \\
\cmidrule(lr){2-3}\cmidrule(lr){4-5}
& $L_{\mathrm{w}}$ & $P_{\mathrm{crit}}$
& $L_{\mathrm{w}}$ & $P_{\mathrm{crit}}$ \\
\midrule
Our solution    & 2.82 & 0.53 & 4.10 & 0.51 \\
MCTS-Rollout   & 2.73 & 0.50 & 3.97 & 0.51 \\
ActorCritic-RL & 2.02 & 0.20 & 3.24 & 0.28 \\
DQN-RL         & 1.94 & 0.27 & 2.87 & 0.27 \\
\bottomrule
\end{tabularx}
\end{table}

\begin{table}[t]
\centering
\caption{IEEE 69-bus restoration results.}
\label{tab:ieee69_results}
\footnotesize
\setlength{\tabcolsep}{4pt}
\renewcommand{\arraystretch}{1.10}
\begin{tabularx}{\columnwidth}{@{}Xcccc@{}}
\toprule
\multirow{2}{*}{Method} &
\multicolumn{2}{c}{$T_{\max}=10$} &
\multicolumn{2}{c}{$T_{\max}=15$} \\
\cmidrule(lr){2-3}\cmidrule(lr){4-5}
& $L_{\mathrm{w}}$ & $P_{\mathrm{crit}}$
& $L_{\mathrm{w}}$ & $P_{\mathrm{crit}}$ \\
\midrule
Our solution    & 3.20 & 0.93 & 4.28 & 1.19 \\
MCTS-Rollout   & 3.01 & 0.87 & 4.01 & 1.12 \\
ActorCritic-RL & 2.39 & 0.46 & 2.75 & 0.28 \\
DQN-RL         & 2.24 & 0.30 & 2.07 & 0.20 \\
\bottomrule
\end{tabularx}
\end{table}

\begin{table}[t]
\centering
\caption{IEEE 118-bus restoration results.}
\label{tab:ieee118_results}
\footnotesize
\setlength{\tabcolsep}{4pt}
\renewcommand{\arraystretch}{1.10}
\begin{tabularx}{\columnwidth}{@{}Xcccc@{}}
\toprule
\multirow{2}{*}{Method} &
\multicolumn{2}{c}{$T_{\max}=10$} &
\multicolumn{2}{c}{$T_{\max}=15$} \\
\cmidrule(lr){2-3}\cmidrule(lr){4-5}
& $L_{\mathrm{w}}$ & $P_{\mathrm{crit}}$
& $L_{\mathrm{w}}$ & $P_{\mathrm{crit}}$ \\
\midrule
Our solution    & 3.23 & 0.67 & 4.63 & 0.94 \\
MCTS-Rollout   & 2.47 & 0.61 & 4.47 & 0.91 \\
ActorCritic-RL & 2.18 & 0.41 & 3.48 & 0.58 \\
DQN-RL         & 1.98 & 0.36 & 3.21 & 0.47 \\
\bottomrule
\end{tabularx}
\end{table}
Tables~\ref{tab:ieee33_results}--\ref{tab:ieee118_results} report restoration
performance across the IEEE 33-bus, 69-bus, and 118-bus test systems under
$T_{\max}=10$ and $T_{\max}=15$. Here, $L_{\mathrm{w}}$ denotes the weighted
restored load, and $P_{\mathrm{crit}}$ denotes the restored critical-load. Across all six settings, our solution achieves the highest $L_{\mathrm{w}}$.
This result is important because $L_{\mathrm{w}}$ reflects the
priority-aware restoration objective, rather than only measuring the total
amount of restored load.

Our solution also provides strong critical-load restoration. It achieves the
highest $P_{\mathrm{crit}}$ value in five of the six settings and matches
MCTS-Rollout in the remaining setting. In the tied critical-load case,
namely the IEEE 33-bus system at $T_{\max}=15$, our solution still achieves
higher $L_{\mathrm{w}}$. This indicates that, after restoring a comparable
amount of critical load, our solution restores a stronger overall combination
of critical and priority-weighted noncritical loads.

Compared with DQN-RL and ActorCritic-RL, our solution achieves higher
$L_{\mathrm{w}}$ and stronger $P_{\mathrm{crit}}$ across all the tested systems.
The learning-based baselines improve with a longer restoration horizon in some cases, but they remain below our solution and MCTS-Rollout. This suggests
that the proposed framework can generate effective restoration strategies in
held-out outage scenarios without relying on learned value estimates
from the training scenarios.

The comparison with MCTS-Rollout is also important because MCTS performs
lookahead search using simulated restoration actions. Our solution remains
competitive with or better than MCTS-Rollout across the tested settings,
while also providing structured LLM-based planning, memory-guided reasoning,
and operator-facing explanations. These results support the proposed design,
where LLM agents generate and select multi-step restoration plans over
shortlisted feasible actions, while deterministic AC power-flow evaluation
checks feasibility before execution.

\subsection{LLM-Agent Structured-Output Reliability}
\label{subsec:llm_output_reliability}

To evaluate the reliability of the LLM-agents layer, we record whether each
agent response can be parsed as a valid structured JSON output. If an agent
response is missing, invalid, or cannot be parsed, the deterministic fallback
mechanism is activated. This analysis therefore measures both structured
output reliability and the frequency of fallback behavior activation.

\begin{table}[t]
\centering
\caption{LLM-agent structured-output reliability for the IEEE 118-bus system.}
\label{tab:llm_structured_output_ablation}
\begin{tabular}{lccc}
\toprule
Agent & Horizon & Invalid Outputs & Success Rate \\
\midrule
Planner Guidance Agent   & $T_{\max}=10$ & 0 & 100\% \\
Strategic Planning Agent & $T_{\max}=10$ & 0 & 100\% \\
Decision Agent          & $T_{\max}=10$ & 0 & 100\% \\
Explainability Agent    & $T_{\max}=10$ & 0 & 100\% \\
\midrule
Planner Guidance Agent   & $T_{\max}=15$ & 0 & 100\% \\
Strategic Planning Agent & $T_{\max}=15$ & 0 & 100\% \\
Decision Agent          & $T_{\max}=15$ & 0 & 100\% \\
Explainability Agent    & $T_{\max}=15$ & 0 & 100\% \\
\bottomrule
\end{tabular}
\end{table}

Table~\ref{tab:llm_structured_output_ablation} shows that all four LLM
agents returned valid structured outputs for both restoration horizons in
the IEEE 118-bus system. No invalid JSON responses were observed, and no
structured-output fallback was triggered. This result suggests that the
prompting, JSON extraction, and validation pipeline was reliable in the
tested cases.

\subsection{Explainability Case Study}

Beyond restoration performance, the proposed framework is designed to
provide operator-facing explanations for its restoration decisions. To
illustrate this capability, this subsection presents a representative
decision-level explanation generated for the IEEE 33-bus
test system in table~\ref{tab:explainability_case_study}.

In this case, our solution selected the action
\textbf{serve\_load:5} as the first action of the selected plan. The
selected plan was
\textbf{serve\_load:5} $\rightarrow$ \textbf{serve\_load:6}
$\rightarrow$ \textbf{serve\_load:8}. The deterministic evaluation showed
that this plan was feasible, produced no voltage
violations or line overloads, restored one critical load, and achieved the
highest weighted load score among the candidate plans.

\begin{table}[!t]
\centering
\caption{Representative explanation generated by our solution for an IEEE
33-bus restoration decision.}
\label{tab:explainability_case_study}
\scriptsize
\setlength{\tabcolsep}{2pt}
\renewcommand{\arraystretch}{1.08}
\begin{tabular}{p{0.27\columnwidth} p{0.66\columnwidth}}
\toprule
Item & Explanation Artifact \\
\midrule
Selected action &
\textbf{serve\_load:5} \\

Selected plan &
\textbf{serve\_load:5} $\rightarrow$
\textbf{serve\_load:6} $\rightarrow$
\textbf{serve\_load:8} \\

Planner guidance &
Preferred action types: \textbf{serve\_load} and \textbf{start\_dg};
avoided action types: \textbf{repair\_line} and \textbf{switch\_tie};
priority bus focus: buses 5 and 18. \\

Evaluation outcome &
The selected plan was feasible, produced no voltage violations or line
overloads, restored one critical load, and achieved strong weighted
restoration among the candidate plans. \\

Reason for selection &
The plan was selected because it restored a critical load, achieved strong
weighted restored load among the candidates, and used immediately reachable
load-restoration actions without requiring additional setup. \\

Reason alternatives were rejected &
Alternative plans were feasible, but they had lower weighted restored load
or did not restore critical load compared with the selected plan. \\
\bottomrule
\end{tabular}
\end{table}

This case study shows how the proposed framework provides a transparent
decision trace. The explanation connects the recommended action to
measurable restoration quantities, including critical-load
restoration, weighted restored load, feasibility status, voltage violations, and
line overloads. Therefore, the proposed framework does
not only recommend a restoration action; it also provides a structured
rationale that can support operator review and decision trust.

\subsection{Discussion}

Overall, the results show that our solution improves priority-aware
restoration performance while maintaining deterministic feasibility checks.
The gains over DQN-RL and ActorCritic-RL suggest that LLM-based multi-step
planning can be more effective than learned value estimates in the held-out
outage scenarios. Compared with MCTS-Rollout, our solution remains competitive
or better while also providing structured explanations for operator review.

\section{Agentic Decomposition Analysis}
\label{sec:agentic_decomposition}

The previous results compare our solution with reinforcement learning and
planning baselines. However, those comparisons do not isolate whether the
performance gain comes from using an LLM in general or from the proposed
multi-agent decomposition. To examine this design choice, this section
compares our solution against a SingleLLMAgent baseline.

In this analysis, the key difference is the
structure of LLM reasoning. Our solution decomposes the restoration process
across specialized agents for planner guidance, strategic plan generation,
decision arbitration, memory-based feedback, and explanation generation. In
contrast, SingleLLMAgent performs all these within one LLM call. Therefore, this comparison
evaluates whether the proposed agentic decomposition provides value beyond a
monolithic LLM decision structure.

The prompt used for the SingleLLMAgent baseline is summarized as follows:

\begin{agentprompt}
You are a single LLM restoration agent. You must do all reasoning tasks by
yourself: (1) provide rollout guidance, (2) generate candidate restoration
plans, (3) select the best plan index, and (4) explain the selected decision.
Return strict JSON containing \textbf{planner\_guidance}, \textbf{plans},
\textbf{selected\_plan\_index}, and \textbf{explanation}. Use only the
provided candidate actions. Do not invent actions. Generate four to six
plans when possible. Each plan should contain three to six actions when
possible. The selected plan index must refer to one of the generated plans.
Prefer feasible, high-value, critical-load-restoring plans. The planning
horizon must be an integer between one and six. Return only valid JSON and
no extra text.
\end{agentprompt}

This baseline is intentionally strong because it is not an unconstrained LLM
controller. It receives the same shortlisted candidate actions as
our agentic structure, and its generated plans are still validated using
deterministic physics-aware evaluation before execution. Thus, the
comparison focuses on whether separating the LLM reasoning process into
specialized agent roles is more effective than asking a single LLM agent to
perform all reasoning tasks simultaneously.

\begin{strip}
\centering
\captionof{table}{Comparison between our solution and SingleLLMAgent. Token counts are reported in thousands.}
\label{tab:agentic_vs_single_llm}
\vspace{0.5em}
\scriptsize
\setlength{\tabcolsep}{3pt}
\renewcommand{\arraystretch}{1.05}
\begin{tabular}{c c l c c c c c c}
\toprule
System & $T_{\max}$ & Method & Weighted Load & Critical Load & Runtime (min) & Input Tok. (k) & Output Tok. (k) & Total Tok. (k) \\
\midrule
33  & 10 & SingleLLMAgent & 2.55 & 0.38 & 6.626 & 18.367 & 7.337 & 25.744 \\
33  & 10 & Our solution    & \textbf{2.82} & \textbf{0.53} & 9.009 & 56.425 & 12.225 & 68.650 \\
\midrule
33  & 15 & SingleLLMAgent & 3.88 & 0.48 & 13.000 & 17.128 & 11.154 & 28.281 \\
33  & 15 & Our solution    & \textbf{4.10} & \textbf{0.51} & 16.070 & 85.601 & 18.140 & 103.741 \\
\midrule
69  & 10 & SingleLLMAgent & 2.38 & 0.36 & 14.440 & 12.047 & 6.972 & 19.019 \\
69  & 10 & Our solution    & \textbf{3.20} & \textbf{0.93} & 19.960 & 63.631 & 11.756 & 75.387 \\
\midrule
69  & 15 & SingleLLMAgent & 3.51 & 0.52 & 27.780 & 17.980 & 10.448 & 28.428 \\
69  & 15 & Our solution    & \textbf{4.28} & \textbf{1.19} & 36.760 & 97.391 & 17.671 & 115.062 \\
\midrule
118 & 10 & SingleLLMAgent & 2.26 & 0.28 & 24.470 & 12.977 & 6.723 & 19.700 \\
118 & 10 & Our solution    & \textbf{3.23} & \textbf{0.67} & 26.480 & 67.023 & 11.131 & 78.154 \\
\midrule
118 & 15 & SingleLLMAgent & 3.32 & 0.44 & 41.210 & 20.593 & 10.538 & 31.131 \\
118 & 15 & Our solution    & \textbf{4.63} & \textbf{0.94} & 48.070 & 109.247 & 17.592 & 126.839 \\
\bottomrule
\end{tabular}
\vspace{0.5em}
\end{strip}

Table~\ref{tab:agentic_vs_single_llm} shows that our solution consistently
outperforms SingleLLMAgent in both weighted restored load and restored
critical-load across all tested systems and restoration horizons. Since
both methods use the same restoration environment, shortlisted candidate
actions, memory summary, and deterministic physics-aware plan evaluation,
the improvement is associated with the decomposition of LLM reasoning into
specialized agent roles rather than merely with using an LLM.

Across the six settings, our solution improves weighted restored load by
approximately 25.8\% on average relative to SingleLLMAgent. The relative
weighted-load gains are 10.6\% and 5.7\% for the IEEE 33-bus system,
34.5\% and 21.9\% for the IEEE 69-bus system, and 42.9\% and 39.5\% for
the IEEE 118-bus system under $T_{\max}=10$ and $T_{\max}=15$,
respectively. These results indicate that the benefit of agentic
decomposition becomes more visible as the feeder size and restoration search
space increase.

Our solution also improves restored critical-load in all six settings.
The largest gains occur in the IEEE 69-bus and IEEE 118-bus systems, where
the restoration problem contains more possible actions and longer
reachability dependencies. This suggests that separating planner guidance,
strategic plan generation, decision arbitration, memory feedback, and
explanation can improve critical-load prioritization compared with requiring
one LLM call to perform all reasoning tasks jointly.

The improvement comes with additional computational cost. Our solution uses
more runtime and more tokens because multiple LLM agents are queried during
each restoration step. On average, our solution uses approximately 94.64k
tokens per scenario, including 79.89k input tokens and 14.75k output tokens,
compared with 25.38k tokens for SingleLLMAgent, including 16.52k input tokens
and 8.86k output tokens. Using the listed OpenAI API pricing for GPT-5.4 at the time of this study
(\$2.50 per million input tokens and \$15.00 per million output tokens), this corresponds to an
estimated cost of approximately \$0.421 per scenario for our solution and
\$0.174 per scenario for SingleLLMAgent. Although the cost is higher than SingleLLMAgent, it is still low and practical. 
The average runtime increases from 21.25 minutes to 26.06 minutes.
Therefore, the multi-agent design provides stronger restoration performance
and critical-load prioritization, at moderately increased overhead. This performance-cost tradeoff is reasonable for operator-facing restoration
decision support, where decision quality, feasibility, and explanation are
important.
\vspace{-0.3em}

\section{Conclusion}

This paper presents an explainable agentic AI framework for distribution
system load restoration. The framework combines feasible-action generation,
topology-aware action shortlisting, LLM-based multi-step
planning, deterministic AC power-flow evaluation, memory feedback, and
structured explanation. Evaluation on IEEE 33-bus, 69-bus, and 118-bus test
systems shows that the proposed framework improves priority-weighted load
restoration and critical-load recovery compared with reinforcement learning, and
MCTS baselines. The results also show that
decomposing restoration reasoning into specialized LLM agents improves
performance over a monolithic LLM agent, although it increases token usage
and runtime. Overall, the proposed framework provides an effective physics-validated
and explainable decision-support approach for distribution system
restoration at low cost.

\section*{Acknowledgement}
This material is based upon work supported in part by the Department of Energy under Award Number DE-SC0026743.

\FloatBarrier

\bibliographystyle{IEEEtran}
\bibliography{references}

\end{document}